\def\DT{D_{\rm T}}
\def\vT{v_{\rm T}}
\documentstyle[multicol,aps,epsf]{revtex} 
\begin{document}
 
\title{Generalized von Smoluchowski model of reaction rates, \\ 
with reacting particles and a mobile trap}
\author{Aleksandar Donev, Jeff Rockwell, Daniel ben-Avraham\footnote
  {{\bf e-mail:} qd00@clarkson.edu}}

\address{Physics Department, and Clarkson Institute for Statistical
Physics (CISP), \\ Clarkson University, Potsdam, NY 13699-5820}
\maketitle

\begin{abstract} 
We study diffusion-limited coalescence, $A+A\rightleftharpoons A$, in
one dimension, in the presence of a {\it diffusing\/} trap.   The system may
be regarded as a generalization of von~Smoluchowski's model for
reaction rates, in that: (a) it includes reactions between the particles
surrounding the trap, and (b) the trap is mobile---both considerations
which render the model more physically relevant.  As seen from the trap's
frame of reference, the motion of the particles is highly correlated, because
of the motion of the trap.  An exact description of the long-time asymptotic
limit is found using the IPDF method, and exploiting a ``shielding" property 
of reversible coalescence that was discovered recently.  In the case where the
trap also acts as a source---giving birth to particles---the shielding property
breaks down, but we find an ``equivalence principle": Trapping and diffusion of
the trap may be compensated by an appropriate rate of birth, such that the
steady state of the system is identical with the equilibrium state in the
absence of a trap.

\end{abstract}

\section{Introduction}
\label{sec:intro}

Non-equilibrium kinetics of diffusion-limited reactions has been the subject
of much recent
interest~\cite{vanKampen,Haken,Nicolis,Ligget,reaction-reviews,JStatPhys}.  
While equilibrium systems can be completely analyzed by means of standard
thermodynamics methods, and reaction-limited processes are well described by
classical rate equations~\cite{Laidler,Benson}, there exist no such general
approaches to the problem of non-equilibrium, diffusion-limited reactions. 

A fundamental model for the reaction rate in diffusion-limited processes had
been presented by von~Smoluchowski~\cite{Smoluchowski}.  In this model, an ideal
spherical trap is surrounded by a swarm of Brownian particles.  The rate of
absorption of particles into the spherical trap models the reaction rate.  The
Smoluchowski model is limited in two important respects: (a) the particles
react with the spherical trap but {\it not\/} with each other, and (b) the
trap itself does not diffuse, but remains static at the origin.  Both
limitations are unphysical: In real reaction processes all particles interact
(and may react) with each other, and {\it all\/} particles are mobile.  

Several attempts have been made to remove the restriction of an immobile
trap~\cite{Schoonover,Weiss,Wio}.  The problem is complicated by the
apparent correlations in the motion of the surrounding
particles: With a mobile trap, the motion of the surrounding
particles is highly correlated, since a step of the trap to the left, say, in
the lab frame of reference, results in an apparent step to the right of {\it
all\/} the surrounding particles in unison, in the trap's frame of reference. 
Results are restricted to empirical formulas inspired by numerical
simulations~\cite{Schoonover}, or to a number of special cases (immobile
particles~\cite{Weiss}; short times~\cite{Wio}).   As regards reactions between
the surrounding particles, these could hardly be considered, other than
numerically, because few models of diffusion-limited kinetics yield themselves
to exact analysis.  In fact, diffusion-limited coalescence ($A+A\to A$) and
annihilation ($A+A\to 0$) in one dimension alone account for most of the known
exact results~\cite{Coalescence,Doering,Krebs,Henkel,Simon,Droz,Bramson,Torney,%
Spouge,Takayasu,Privman}.

Recently, we have studied reversible coalescence, $A+A\rightleftharpoons A$,
on the line and in the presence of a {\it static\/} trap~\cite{StaticTrap}.  An
exact analysis is possible with the method of interparticle distribution
functions (IPDF)~\cite{DbArevs}.  We have found a remarkable property of
``shielding": The particle nearest to the trap effectively shields the
remaining particles from the trap.  The steady state of the system is uniquely
characterized by the distance of the nearest particle to the trap---all other
particles remain distributed exactly as in the equilibrium state of the system
in the absence of a trap.   This shielding property persists even in the
presence of a bias field (convection, or drift).

In this paper we consider reversible coalescence with a {\it mobile\/} trap
which diffuses with a diffusion constant $\DT$, not necessarily equal to the
diffusion constant of the surrounding particles, $D$.  The problem may still
be formulated with the IPDF method, in spite of the correlations induced by the
motion of the trap.  The shielding property of the nearest particle to the trap
also remains in effect, and it enables us to find a complete exact
description of the distribution of particles in the long-time asymptotic
limit.  We find that relative to the trap the surrounding particles
remain at equilibrium, but the gap to the nearest particle is proportional to
$D+\DT$.  

An interesting generalization is to the case when the backward process $A\to
A+A$ is not limited to the particles alone, but the trap too may act as a
source. Again, the rate of generation of particles from the trap, $\vT$, need
not coincide with $v$, the production rate from the reverse birth reaction.  We
find that if
$\vT>0$ the shielding property breaks down, and we are unable to derive a
complete exact description of the system.  However, if the birth rate is
$\vT=v(1+\DT/D)$, the effect of the trap is nullified: The particles remain
distributed as in equilibrium, as if the trap were not present.

The remainder of this paper is organized as follows. In
section~\ref{sec:model} we review the coalescence process and the IPDF
method used for its analysis. 
The model with a mobile trap is considered in 
section~\ref{sec:mobile_trap}.  In section~\ref{sec:trap-source}, we generalize
to the case where the trap may also act as a source.  We conclude with a
summary and discussion, in section~\ref{sec:discussion}.

\section{Coalescence and the IPDF method}
\label{sec:model}
Our model \cite{Coalescence,Doering,DbArevs} is defined on a one-dimensional
lattice of lattice spacing
$a$.  Each site is in one of two states: occupied by a particle $A$
($\bullet$), or empty ($\circ$). Particles hop randomly to the nearest neighbor
site to their right or left, at rate $D/a^2$.  Thus, in the
continuum limit of
$a\to 0$ the particles undergo diffusion with a diffusion constant $D$.  A
particle may give birth to an additional particle, into a nearest neighbor
site, at rate
$v/a$ (on either side of the particle)\footnote{Our notation here differs from
previous work: we take the birth rate to be $v/a$ rather than $v/2a$, to
achieve a more aesthetic form of the final result.}.  If hopping or birth
occurs into a site which is already occupied, the target site remains
occupied.  The last rule means that coalescence,
$A+A\to A$, takes place {\em immediately\/} upon encounter of any two
particles.  Thus, the system models the
diffusion-limited reaction process
\begin{equation}
A+A\rightleftharpoons A\;. 
\end{equation}
The dynamical rules of the model are illustrated in Fig.~1.
 
%%% FIGURE 1
\begin{figure}
\centerline{\epsfxsize=8cm \epsfbox{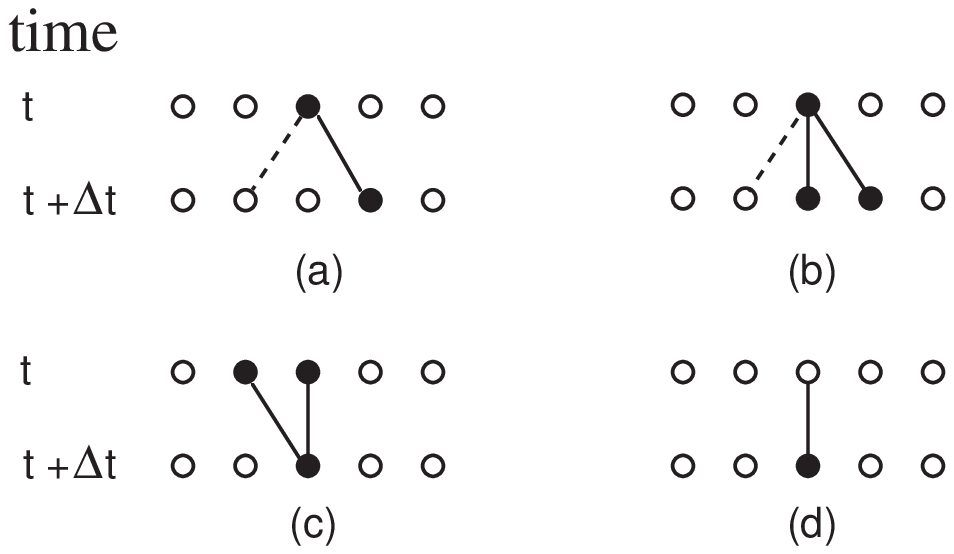}}
\noindent
{\small {\bf Figure~1} 
Reaction rules: (a) diffusion; (b)
birth; and coalescence, (c) following diffusion, and (d) following a birth
event.  The broken lines in (a) and (b) indicate alternative target sites.}
\end{figure} 
%%%

An exact treatment of the problem is possible through the method of Empty
Intervals, known also as the method of Inter-Particle Distribution Functions
(IPDF)~\cite{DbArevs}.  The key concept is
$E_{n,m}(t)$---the probability that sites $n,n+1,\cdots,m$ are empty at time
$t$.  The probability that site $n$ is occupied is 
\begin{equation}
\label{disc-conc}
{\rm Prob}({\rm site\ }n{\rm\ is\ occupied})
\equiv{\rm Prob}\big(\stackrel{n}{\bullet}\big)=1-E_{n,n}\;.
\end{equation}
The event that sites $n$ through $m$ are empty (prob. $E_{n,m}$) consists of two
cases: site $m+1$ is also empty (prob. $E_{n,m+1}$), or it is occupied.  Thus,
the probability that sites $n$ through $m$ are empty but site
$m+1$ is occupied, is 
\begin{equation}
\label{edge1}
{\rm Prob}(\stackrel{n}{\circ}\cdots\stackrel{m}{\circ}\bullet)
=E_{n,m}-E_{n,m+1}\;,  
\end{equation}
and likewise,
\begin{equation}
\label{edge2}
{\rm Prob}(\bullet\stackrel{n}{\circ}\cdots\stackrel{m}{\circ})
=E_{n,m}-E_{n-1,m}\;.  
\end{equation}

With this in mind, one can write down
a rate equation for the evolution of the empty interval probabilities:
\begin{eqnarray}
\label{mastereq}
{\partial E_{n,m}\over\partial t}
   = &&{D\over a^2}(E_{n,m-1}-E_{n,m})\nonumber\\
   - &&{D\over a^2}(E_{n,m}-E_{n,m+1})\nonumber\\
   + &&{D\over a^2}(E_{n+1,m}-E_{n,m})\nonumber\\
   - &&{D\over a^2}(E_{n,m}-E_{n-1,m})\nonumber\\
   - &&{v\over a}[(E_{n,m}-E_{n,m+1})+(E_{n,m}-E_{n-1,m})]\;.
\end{eqnarray}
For example, the first term on the r.h.s. of Eq.~(\ref{mastereq}) accounts for
the increase in $E_{n,m}$ when the particle at the right edge of
$\,\stackrel{n}{\circ}\cdots\circ\!\stackrel{m}\bullet\,$ hops to the right 
and the sites $n,\dots,m$ become empty; the second term denotes the decrease
in $E_{n,m}$ when a particle at $m+1$ hops to the left into the empty interval
$n,\dots,m$, and so on. 

Eq.~(\ref{mastereq}) is valid for
$m>n$.  The special case of
$m=n$ corresponds to $E_{n,n}$---the probability that site $n$ is empty.  It
is described by the equation
\begin{eqnarray}
\label{mastereq_n,n}
{\partial E_{n,n}\over\partial t}
   = &&{D\over a^2}(1-E_{n,n})\nonumber\\
   - &&{D\over a^2}(E_{n,n}-E_{n,n+1})\nonumber\\
   + &&{D\over a^2}(1-E_{n,n})\nonumber\\
   - &&{D\over a^2}(E_{n,n}-E_{n-1,n})\nonumber\\
   - &&{v\over a}[(E_{n,n}-E_{n,n+1})+(E_{n,n}-E_{n-1,n})]\;.
\end{eqnarray}
Comparison with Eq.~(\ref{mastereq}) yields the boundary condition:
$E_{n,n-1}=1$.  The fact that the $\{E_{n,m}\}$ represent {\em probabilities\/}
implies the additional condition that $E_{n,m}\geq 0$.  Finally, if the system
is not empty then $\lim_{{n\to-\infty\atop m\to+\infty}}E_{n,m}= 0$.

In many applications, it is simpler to pass to the continuum limit.  We write
$x=na$ and $y=ma$, and replace $E_{n,m}(t)$ with $E(x,y,t)$.  Letting $a\to 0$,
Eq.~(\ref{mastereq}) becomes
\begin{equation}
\label{eqE}
{\partial \over\partial t}E=D({\partial^2\over\partial x^2}
  +{\partial^2\over\partial y^2})E - v({\partial\over\partial x}
    -{\partial\over\partial y})E\;,
\end{equation}
with the boundary conditions, 
\begin{eqnarray}
\label{BC1}
E(x,x,t) &=& 1\;,\\
\label{BCpositive}
E(x,y,t) &\geq& 0\;,\\
\label{BC0}
\lim_{{x\to-\infty\atop y\to+\infty}}E(x,y,t) &=& 0\;.
\end{eqnarray}
The concentration becomes
\begin{equation}
\label{conc}
\rho(x,t)=-{\partial \over\partial y}E(x,y,t)|_{y=x}\;,
\end{equation}
and one can also show that the conditional joint probability
for having particles at
$x$ and $y$ but none in between, is 
\begin{equation}
\label{P2}
P_2(x,y,t)=-{\partial^2\over\partial x\,\partial y}E(x,y,t)\;.
\end{equation}
{}From $P_2$ one obtains the ``forward" (and also ``backward")
IPDF---the probability that given a particle at
$x$ ($y$) the next nearest particle to its right (left) is at $y$
($x$): 
\begin{equation}
\label{IPDFs}
p_{\rm f}(x,y,t)=\rho(x,t)^{-1}P_2(x,y,t)\;; \qquad 
p_{\rm b}(x,y,t)=\rho(y,t)^{-1}P_2(x,y,t)\;.
\end{equation}

The IPDF method can also handle multiple-point correlation
functions~\cite{Doering}.  Let
$E_n(x_1,y_1,x_2,y_2,\dots,x_n,y_n,t)$ be the joint probability that the
intervals $[x_i,y_i]$ ($i=1,2,\dots,n)$ are empty at time $t$.  The
intervals are non-overlapping, and ordered: $x_1<y_1<\cdots<x_n<y_n$. Then,
the $n$-point correlation function (the probability of finding particles at
$x_1,x_2,\dots,x_n$ at time $t$) is given by
\begin{equation}
\label{rho_n}
\rho_n(x_1,\dots,x_n,t)=
(-1)^n{\partial^n\over\partial y_1\cdots\partial y_n}
E_n(x_1,y_1,\dots,x_n,y_n,t)|_{y_1=x_1,\dots,y_n=x_n}\;.
\end{equation}
For reversible coalescence, the $E_n$ satisfy the partial differential equation:
\begin{eqnarray}
\label{dEn/dt}
{\partial\over\partial t} E_n(x_1,y_1,\dots,x_n,y_n,t)
&&=D({\partial^2\over\partial x_1^2} +{\partial^2\over\partial y_1^2} + \cdots
+{\partial^2\over\partial x_n^2} + {\partial^2\over\partial
 y_n^2})E_n\nonumber\\
&&-v[({\partial\over\partial x_1}-{\partial\over\partial y_1}) + \cdots
+({\partial\over\partial x_n} - {\partial\over\partial y_n})]E_n\;,
\end{eqnarray}
with the boundary conditions
\begin{equation}
\label{bc:En.1}
\lim_{x_i\uparrow y_i{\rm\ or\ }y_i\downarrow x_i}
  E_n(x_1,y_1,\dots,x_n,y_n,t)=
  E_{n-1}(x_1,y_1,\dots,\not{\!x_i},\not{\!y_i},\dots,x_n,y_n,t)\;,
\end{equation}
and
\begin{equation}
\label{bc:En.2}
\lim_{y_i\uparrow x_{i+1}{\rm\ or\ }x_{i+1}\downarrow y_i}
 E_n(x_1,y_1,\dots,x_n,y_n,t)=
 E_{n-1}(x_1,y_1,\dots,\not{\!y_i},\not{\!x_{i+1}},\dots,x_n,y_n,t)\;.
\end{equation}
For convenience, we use the notation that crossed out arguments (e.g.
${\not{\!x_i}}$) have been removed.  The $E_n$ are tied together
in an hierarchical fashion through the boundary conditions~(\ref{bc:En.1})
and (\ref{bc:En.2}):  one must know $E_{n-1}$ in order to compute $E_n$.

As a trivial example, consider the homogeneous steady state of reversible
coalescence.  This is in fact an {\it equilibrium\/} state, which satisfies
detailed balance.  The particles are simply distributed completely randomly---a
state which maximizes their entropy.  One obtains
\begin{equation}
\label{En.eq}
E_{n,{\rm eq}}=\exp\{-\gamma[(y_1-x_1)+\cdots+(y_n-x_n)]\}\;,
\end{equation}
and 
\begin{equation}
\rho_{n,{\rm eq}}(x_1,x_2,\dots,x_n)=\gamma^n\;, 
\end{equation}
where $\gamma\equiv v/D$ is the
particle concentration at equilibrium.

\section{Coalescence with a mobile trap}
\label{sec:mobile_trap}
We now consider the coalescence model but with a trap which diffuses with a
diffusion constant $\DT$:  In the discrete representation, the trap hops to
its right or left at rate $\DT/a^2$.  A particle that hops into the trap
is irreversibly captured by it.  Similarly, when the trap hops onto an
occupied site it captures the particle in that site.  It is convenient to
analyze the system in the trap's frame of reference.  In this view, the trap
remains static at a site which we choose to be the origin; $n=0$.  When the
trap does not move (in the lab frame of reference) the changes to
$E_{n,m}$ are described by Eq.~(\ref{mastereq}).  A motion of the trap
is perceived as a coherent opposite motion of the particles in the trap's
reference frame.  Thus, the changes to $E_{n,m}$ due to the motion of the
trap are:
\begin{eqnarray}
\label{Enm.trap}
{\partial\over\partial t}(E_{n,m})_{\rm trap}
   =   {\DT\over a^2}[
  &&   (E_{n+1,m+1}-E_{n,m+1}) 
     + (E_{n-1,m-1}-E_{n-1,m})\nonumber\\
  && - (E_{n,m}-E_{n-1,m}) 
     - (E_{n,m}-E_{n,m+1})]\;.
\end{eqnarray}
For example, the first term on the r.h.s. denotes the possibility that site
$n+1$ is occupied while the subsequent sites $n+2,n+3,\dots,m+1$ are empty,
and the trap hops to the right: In the trap's frame of reference the particle
at $n+1$ seems to hop to the left, thereby clearing the $[n,m]$-interval. 
Notice that it is important to make sure that site $m+1$ is empty, since
otherwise site $m$ would become occupied as the trap moves to the right.   

Putting together all the terms in~(\ref{mastereq}) and
(\ref{Enm.trap}), and passing to the continuum limit, we obtain
\begin{equation}
\label{eqE.mobile}
{\partial\over\partial t}E=(D+\DT)({\partial^2\over\partial x^2}
  +{\partial^2\over\partial y^2})E 
  + 2\DT{\partial^2\over\partial x\partial y}E
  - v({\partial\over\partial x}-{\partial\over\partial y})E\;,
\end{equation}
which is now valid in the infinite wedge $0<x<y$.
The term with the mixed derivative is special: it arises because
of the correlated motion of the particles in the reference
frame of the trap.

The trap at $n=0$ could be realized by holding that site empty, at all
times.  Thus, it follows that $E_{1,m}=E_{0,m}$, which results in the
boundary condition (in the continuum limit):
\begin{equation}
\label{BC:dE/dx}
{\partial\over\partial x}E(x,y,t)|_{x=0}=0\;.
\end{equation}
In addition, the boundary condition~(\ref{BC0}) becomes
\begin{equation}
\label{BC0.trap}
\lim_{y\to\infty}E(x,y,t) = 0\;,
\end{equation}
while~(\ref{BC1}) and (\ref{BCpositive}) still apply, without change.

We now search for a solution of Eq.~(\ref{eqE.mobile}), with the boundary
conditions~(\ref{BC1}), (\ref{BCpositive}), (\ref{BC:dE/dx}), and
(\ref{BC0.trap}), in the long-time asymptotic limit, $\partial E/\partial
t=0$.  It is simple to find eigenfunctions which obey Eq.~(\ref{BC1}), and
other eigenfunctions which obey Eq.~(\ref{BC:dE/dx}), but we were unable to
devise a systematic method for finding the linear combinations that would
satisfy both conditions simultaneously.  Instead, we offer a solution
based on the newly discovered property of ``shielding" in the coalescence
model~\cite{StaticTrap,Waves}.

In the steady state of the coalescence model with a {\it static\/} trap, it
is found that the particles are distributed randomly, exactly as in the
equilibrium state of the homogeneous, infinite system (end of
section~\ref{sec:model}).  The system is then fully characterized by
$p(z)$---the density distribution function of the distance between the trap
and the nearest particle to the trap, $z$.  The nearest particle effectively
shields the remaining particles from the trap (Fig.~2).  As we show below,
the same shielding effect takes place even when the trap is mobile.
 
%%% FIGURE 2
\begin{figure} 
\centerline{\epsfxsize=8cm \epsfbox{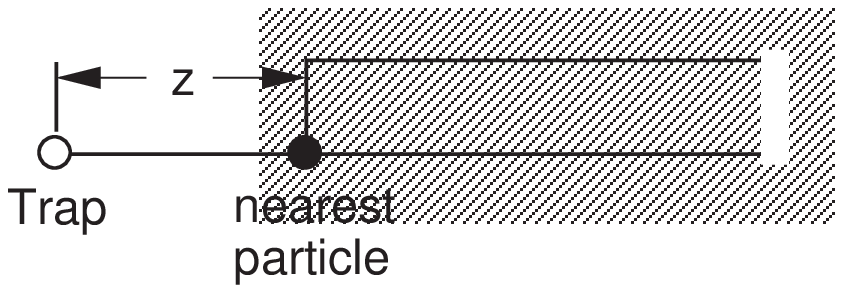}}
\noindent
{\small {\bf Figure~2}. 
Schematic illustration of the shielding effect:  The
particles in the shaded area are distributed randomly and independently from
each other, as in equilibrium.  The gap $z$ between the particles and the trap
follows the probability density distribution $p_0(z)$.}
\end{figure} 
%%%

Assuming that shielding holds, let $E(x,y|z)$ be the conditional probability
that the interval $[x,y]$ is empty, given that the nearest particle to the
trap is at $z$, then:
\begin{equation}
E(x,y|z)=\left\{ \begin{array}{ll}
1, & \mbox{$x<y<z$}\;,\\ 
0, & \mbox{$x<z<y$}\;,\\
e^{-\gamma(y-x)}, & \mbox{$z<x<y$}\;. \end{array}
\right.
\end{equation}
Hence, 
\begin{eqnarray}
\label{Eedge}
E(x,y)&=&\int_0^{\infty}E(x,y|z)p(z)\,dz=\int_y^{\infty} p(z)\,dz 
     + e^{-\gamma(y-x)}\int_0^x p(z)\,dz\nonumber\\
&=& 1-F(y)+e^{-\gamma(y-x)}F(x)\;,
\end{eqnarray}
where in the last equation we introduced the definition
\begin{equation}
\label{defF}
F(z)\equiv\int_0^z p(z')\,dz'\;.
\end{equation}
If the hypothesis of shielding is correct, the particle concentration can be
obtained from Eqs.~(\ref{conc}) and (\ref{Eedge}): 
\begin{equation}
\label{conc.F}
\rho(x)=p(x)+\gamma F(x)\;.
\end{equation}

Substituting $E(x,y)$ from Eq.~(\ref{Eedge}) into Eq.~(\ref{eqE.mobile}),
in the stationary limit, the variables separate:
\begin{equation}
[(D+\DT){\partial^2\over\partial x^2}F(x) 
  + v{\partial\over\partial x}F(x)]e^{\gamma x} =
[(D+\DT){\partial^2\over\partial y^2}F(y) 
  + v{\partial\over\partial y}F(y)]e^{\gamma y}\;,
\end{equation}
and so, one is lead to the conclusion that
\begin{equation}
\label{eqF}
(D+\DT){d^2\over dz^2}F(z)+v{d\over dz}F(z)=C'e^{-\gamma z}\;,
\end{equation}
where $C'$ is a constant.

The general solution of Eq.~(\ref{eqF}) is
\begin{equation}
\label{F.ABC}
F(z)=A+Be^{-\gamma'z}+Ce^{-\gamma z}\;,
\end{equation}
where $A$, $B$, and $C=C'D^2/v^2\DT$ are constants, to be determined from
boundary conditions.  From the definition of $F$, we have;
$F(0)=0$,
$\lim_{z\to\infty}F(z)=1$, and
$F(z)\geq 0$.  The boundary condition due to the presence of the trap
(Eq.~\ref{BC:dE/dx}) translates into $dF/dz|_{z=0}=0$. Thus, we find
\begin{equation}
\label{F}
F(z)=1+{D\over\DT}e^{-\gamma z}-{D+\DT\over\DT}e^{-\gamma'z}\;,
\end{equation}
where $\gamma\,'\equiv v/(D+\DT)$.
It then follows that
\begin{equation}
\label{p}
p(z)={v\over\DT}(e^{-\gamma'z}-e^{-\gamma z})\;.
\end{equation}
{}From $p(z)$ one immediately obtains the average distance between the
trap and the nearest particle:
$(2D+\DT)/v$, as well as the particle concentration in the trap's frame of
reference (Eq.~\ref{conc.F}):
\begin{equation}
\label{rho_ideal}
\rho(x)=\gamma(1-e^{-\gamma'x})\;.
\end{equation}
The last result is similar to the one obtained for a static trap, only that
the width of the depletion zone near the trap is $1/\gamma\,'=(D+\DT)/v$,
instead of $1/\gamma=D/v$.

Our original goal of finding the empty interval probability
has now been achieved.  Using~(\ref{Eedge}) and (\ref{F}), we
get
\begin{equation}
E(x,y)=e^{-\gamma(y-x)}
  +{D+\DT\over\DT}e^{-\gamma'y}[1-e^{-(\gamma-\gamma')(y-x)}]\;.
\end{equation}
This solution can be verified by direct substitution in Eq.~(\ref{eqE.mobile}) 
and in the boundary conditions~(\ref{BC1}), (\ref{BC:dE/dx}), and
(\ref{BC0.trap}).  The fact that we have found a
solution proves that  shielding indeed takes place, even with a mobile trap.
On the other hand, we have merely shown that $E(x,y)$ is consistent with
the shielding assumption.  We now wish to show that the same is true for the
whole hierarchy of $E_n$'s, and hence for all $n$-point correlation functions.

If the particles are distributed as implied by shielding, then,
following a reasoning similar to that which led to Eq.~(\ref{Eedge}), we
should have
\begin{eqnarray}
\label{En}
E_n&&(x_1,y_1,\dots,x_n,y_n)=1-F(y_n)\nonumber\\
&&+e^{-\gamma(y_n-x_n)}[F(x_n)-F(y_{n-1})]\nonumber\\
&&+\cdots+e^{-\gamma\{(y_n-x_n)+\cdots+(y_i-x_i)\}}[F(x_i)-F(y_{i-1})]
                                           \nonumber\\
&&+\cdots+e^{-\gamma\{(y_n-x_n)+\cdots+(y_1-x_1)\}}F(x_1)\;.
\end{eqnarray}
It is easy to confirm that these functions fulfill the boundary 
conditions~(\ref{bc:En.1}) and (\ref{bc:En.2}).
Eq.~(\ref{dEn/dt}) is also satisfied, provided that $F$ satisfies the
same equation as above, Eq.~(\ref{eqF}), with the same boundary conditions. 
That is, the solution found above for $F$, combined with Eq.~(\ref{En}),
solves the problem of the $E_n$.  Indeed, using Eqs.~(\ref{conc.F}),
(\ref{rho_n}), and (\ref{En}), we find the
$n$-point correlation function:
\begin{equation}
\rho_n(x_1,\dots,x_n)=\rho(x_1)\gamma^{n-1}\;,
\end{equation}
exactly as we expect from a system with the shielding property.

\section{Coalescence with a trap-source}
\label{sec:trap-source}
We now wish to consider a further generalization of the trapping problem. 
Suppose that the trap is imperfect, in the sense that it may also act as a
source: the trap gives birth to $A$ particles into the site next to it, at
rate $\vT/a$.  When $\DT=D$ and $\vT=v$ the trap is identical to the rest of
the particles (only that the system is empty to the left of the trap). 
Thus, such a trap-source might also be viewed as a special, tagged particle,
characterized perhaps by a different diffusing constant and back reaction
rate.  The model constitutes a modest first step towards the understanding of
the more realistic situation where the size of  aggregates matters, and
clusters diffuse and give birth at different rates, determined by their
accumulated mass.  A very recent  application is to the
evolution of bacterial colonies living near a patch of nutrients.  Nelson et
al.,~\cite{Nelson,Dahmen} analyze such experiments with a
diffusion-limited coalescence model with a source (modeling the
nutrients) similar to ours.

The evolution equation for empty intervals in the trap-source model is
identical to that of a perfect trap (Eq.~\ref{eqE.mobile}). The birth of
particles from the trap affects only the boundary condition at $x=0$: it is
no longer true that the trap may be realized by simply holding site $n=0$
empty.  To derive the appropriate boundary condition we consider the total
changes to
$E_{n,m}$, which are obtained by putting together Eqs.~(\ref{mastereq}) and
(\ref{Enm.trap}).  The case of
$n=1$ needs to be considered separately, since we do not know what is
$E_{0,m}$.  The changes to
$E_{1,m}$, including birth from the trap, add up to:
\begin{eqnarray}
\label{mastereq_1,m}
{\partial\over\partial t}E_{1,m}
   = &&{D\over a^2}[(E_{1,m-1}-E_{1,m})-(E_{1,m}-E_{1,m+1})
        +(E_{2,m}-E_{1,m})]\nonumber\\
  && +{\DT\over a^2}[(E_{2,m+1}-E_{1,m+1})
        -(E_{1,m}-E_{1,m+1})+(E_{1,m-1}-E_{1,m})]\nonumber\\
  && -{v\over a}(E_{1,m}-E_{1,m+1}) -{\vT\over a}E_{1,m}\;.
\end{eqnarray}
Comparison of this equation with that for general $n$, when $n=1$,
yields the discrete boundary condition:
\begin{equation}
\label{BCtrap-source.disc}
({D\over a^2}+{v\over a})(E_{1,m}-E_{0,m})
  +{\DT\over a^2}(E_{1,m-1}-E_{0,m-1}) = {\vT\over a}E_{1,m}\;.
\end{equation}
Notice that when $\vT=0$ this reduces to $E_{0,m}=E_{1,m}$, as for a perfect
trap.  Passing to the continuum limit, the boundary condition for the
trap-source becomes
\begin{equation}
\label{BCtrap-source}
(D+\DT){\partial\over\partial x}E(x,y,t)|_{x=0}=\vT E(0,y,t)\;.
\end{equation}

We now seek a solution to Eq.~(\ref{eqE.mobile}), in the stationary limit
$\partial E/\partial t=0$, and which satisfies the boundary conditions
(\ref{BC1}), (\ref{BCpositive}), (\ref{BC0.trap}), and (\ref{BCtrap-source}).
Assuming that shielding holds, we follow the same steps as in
section~\ref{sec:mobile_trap} and we arrive at exactly the same result;
$F(z)=A+Be^{-\gamma'z}+Ce^{-\gamma z}$, only that now the boundary condition
$dF/dz|_{z=0}=0$ is replaced by:
\begin{equation}
\label{bcFtrap-source}
(D+\DT)e^{-\gamma z}{d\over dz}F(z)|_{z=0}=\vT[1-F(z)]\;,
\end{equation}
from Eq.~(\ref{BCtrap-source}).
(Again, notice that when $\vT=0$ one recovers the condition
$dF/dz|_{z=0}=0$.)

{}From the boundary condition $F(\infty)=1$, we obtain $A=1$. Furthermore, from
$F(0)=0$ we get $B=-(1+C)$.  Finally, from the boundary condition due to
the trap-source, Eq.~(\ref{bcFtrap-source}), we get
\begin{equation}
\label{bcC}
[{\vT\over v}\gamma\,'e^{(\gamma-\gamma')z}-1](1+C)=
   ({\vT\over v}\gamma\,'-\gamma)C\;.
\end{equation} 
This condition cannot be satisfied for all $z$ generically, and so one must
conclude that {\it shielding does  not take place in the system with a
trap-source\/}.  On the other hand, for the special case that $C=-1$ and
$\gamma\,'\vT/v=\gamma$, Eq.~(\ref{bcC}) is satisfied. 
In this case $F(z)=1-e^{-\gamma z}$, which leads to $E=e^{-\gamma(y-x)}$
and $\rho(x)=\gamma$.  That is, the particles are distributed exactly as in
equilibrium, as if there were no trap!  Thus, there exists a whole class of
states which are equivalent to the equilibrium state of the
infinite, homogeneous system, without a trap.  The equivalent states are
characterized by the relation
\begin{equation}
\label{vTspecial}
\vT=v(1+{\DT\over D})\;.
\end{equation}
For these states the effect of the trap is nullified. A larger diffusivity of
the trap, $\DT$, (i.e., a larger trapping efficiency) is
exactly compensated by an increasing rate of birth $\vT$ from the trap-source.

An interesting case is when $\DT=D$ and $\vT=v$.  Then the
trap is identical to the surrounding particles.  Notice, however, that this
is {\it not\/} a ``special" state (Eq.~\ref{vTspecial} is not satisfied), and
hence we conclude that at the equilibrium state the system seems {\it
inhomogeneous\/} from the point of view of a tagged particle.  To shed
some light on these baffling results, we first point out that since the
equilibrium state is homogeneous it is perceived without change by any moving
observer which does not interact with the particles, including random
walkers.  [Indeed, the equilibrium state $E=e^{-\gamma(y-x)}$ is a steady-state
solution of Eq.~(\ref{eqE.mobile}).] Imagine then an observer diffusing through
the system with diffusion constant $\DT$, and which does not interact with the
particles.  From the point of view of the observer he is static, and
the average concentration of particles is constant and equal to $\gamma$. 
Ignoring the half infinite line to his left, the observer could
interpret crossings of particles from right to left as ``trapping" events,
provided that he also interprets crossings from left to right as ``birth"
events.  The apparent rate of birth (crossings from left to right) would be
$\vT/a=n[(D+\DT)/a^2+v/a]$, where
$n=\gamma a$ is the average number of particles at the site occupied by the
observer.  Passing to the continuum limit, we recover the ``equivalence"
condition, Eq.~(\ref{vTspecial}).  (Alternatively, one could use the
exact discrete result: $\gamma=v/(D+va)$, to obtain the discrete analogue of
the equivalence condition: $\vT=v[1+\DT/(D+va)]$.)

Although
shielding breaks down when the trap acts also as a source, one may still look
for a solution to the problem in more conventional ways.  We were unable to
find an analytic solution; however, the discrete equations can be integrated
numerically, and the particle system may also be simulated on a computer. 
Computer simulations confirm the fact that shielding breaks down when $\vT>0$,
and that the particle concentration beyond the nearest particle to the trap is
no longer as in equilibrium.  In Fig.~3 we show typical results for for
various values of
$\vT$.  As $\vT$ increases, the concentration of particles near the trap
increases, from zero (for $\vT=0$), to $\gamma$ (for the appropriate
``special" rate, Eq.~\ref{vTspecial}), and to concentrations larger than
$\gamma$.  

%%% FIGURE 3
\begin{figure}
\centerline{\epsfxsize=8cm \epsfbox{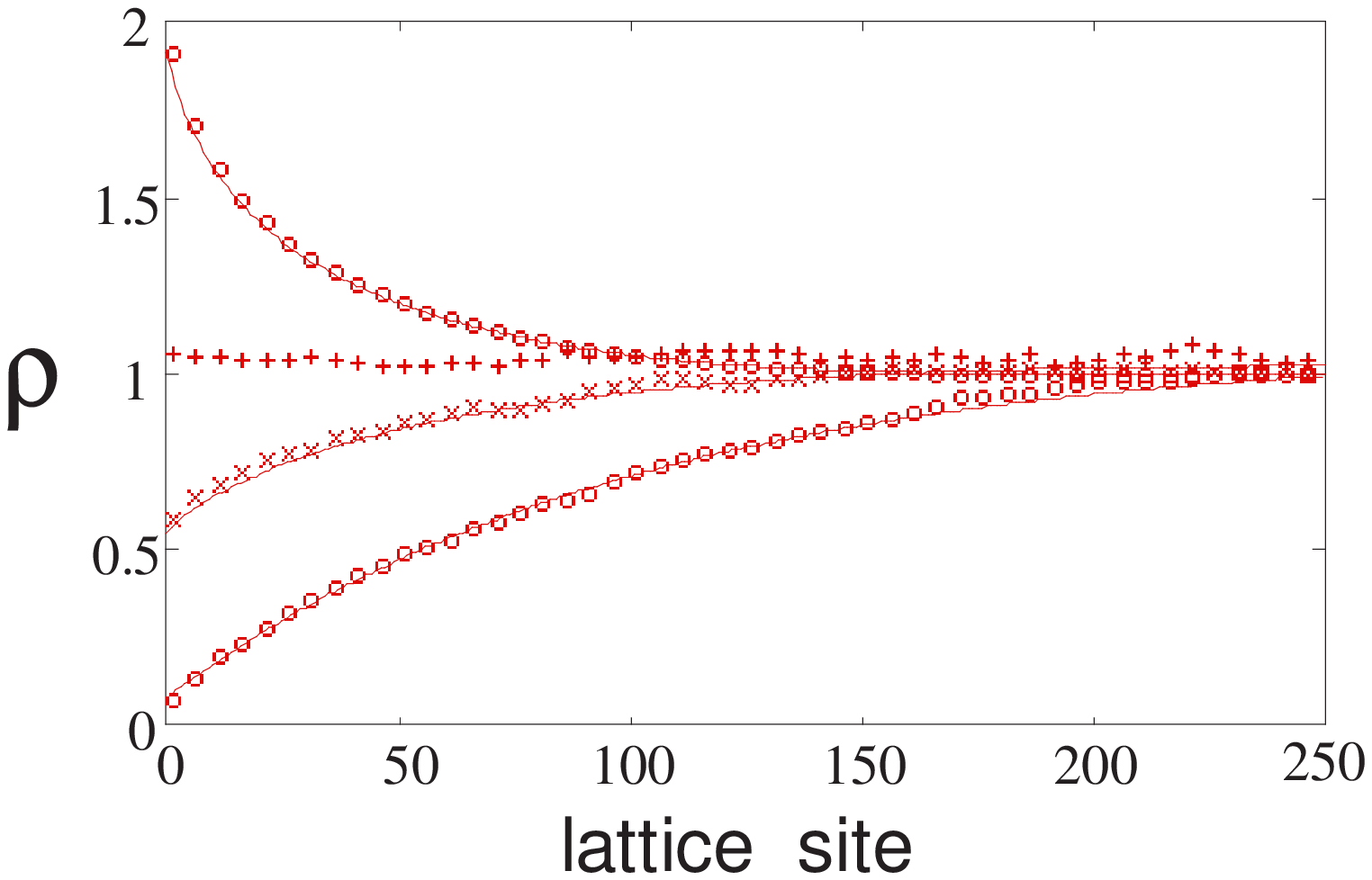}}
\noindent
{\small {\bf Figure~3}. 
Concentration profile with a trap-source: Shown are results
from numerical simulations (symbols) on a lattice of 25,000 sites, averaged
over $a^2/D=10^6$ time steps; as well as results obtained from numerical
integration of the exact discrete equations (solid curves). The cases shown
are for $\vT/v=2$, $1$, $0.5$, and $0$ (top to bottom), with $\DT=0$.}
\end{figure} 
%%%

\section{Discussion}
\label{sec:discussion}

We have studied diffusion-limited coalescence, $A+A\rightleftharpoons A$, 
in the presence of a {\it diffusing\/} trap, and we have found an exact
description of the long-time asymptotic limit, using the IPDF method.
When the trap is perfect, the system displays a ``shielding" property: the
particle nearest to the trap effectively shields the other particles from
the trap.  That is, the particles remain distributed as in the equilibrium
steady-state of the infinite homogeneous system (without a trap), and only the
distance between the trap and the nearest particle is unusual.  This
distance grows linearly with $\DT$---the diffusion coefficient of the trap.

For an
imperfect trap which also acts as a source the shielding property breaks down
and we were unable to find an analytic solution, but the {\it exact\/}
equations can then be solved numerically.   We have found an intriguing
``equivalence principle": all systems with
$\vT=v(1+\DT/D)$ are equivalent to each other.  The trap then seems
invisible and the particles remain distributed as in the homogeneous
equilibrium state.

Our system is a generalization of von Smoluchowski's model for reaction rates;
the particles react with each other, and the trap is mobile.  The
reaction rate equals the rate of influx of particles into the trap.  This
rate is $k=n[(D+\DT)/a^2+v/a]$, where $n$ is the average number of particles
at the site adjacent to the trap. For a perfect trap, we use the result
of Eq.~(\ref{rho_ideal}) to find
$n=a^2(d\rho/dx)_{x=0}=a^2\gamma\gamma'=a^2v^2/D(D+\DT)$, and so, in the
continuum limit ($a\to 0$) we get $k=v^2/D$.  Curiously, the
trapping rate is independent of the diffusivity of the trap, $\DT$.  A
faster trap visits more sites per unit time, but it also depletes its immediate
neighborhood more effectively, and the two effects cancel each other.  For
the case of a trap-source, we have failed to obtain an analytic expression for
$n$, and hence we could compute $k$ only numerically.

There remain several interesting open problems.  We have considered only the
steady state of our model, but the transient is also of interest.  For
perfect traps the shielding property holds at all times (provided that the
initial condition is compatible with it) and one can exploit it to find an
analytic answer.  An important open problem is that of finding a systematic
method for solving the evolution equation for empty intervals.  We were
fortunate to come across a solution which obeys shielding, but shielding does
not always hold, as for example for non-ideal trap-sources.  Indeed, most
remaining open questions concern the model with a trap-source.  An exact
analytic solution for this case is still missing.  We have managed to prove,
however, that the solution could not be of the form of a sum of exponentials
(finite {\it or\/} infinite), other than for the special states, equivalent to
equilibrium.

An interesting question is whether there exist other classes of
equivalence.  That is, are there any states equivalent to each other, but not
to the equilibrium state?  ---We have managed to prove that such states do
not exist, at the level of the empty interval probability.  However, there
remains the possibility that different systems might share the same {\it
concentration\/} profile, in spite of differences in their empty interval
probabilities.  Whether such states exist remains an open challenge.

%%%%%%%%%%%%%%%%%%%%%%%%%%%%%%%%%%%%%%%%%%%%%%%%%%%%%%%%%%%%%%%%%%%%%%%%%%%%%%%%
%% REFERENCES
%%%%%%%%%%%%%%%%%%%%%%%%%%%%%%%%%%%%%%%%%%%%%%%%%%%%%%%%%%%%%%%%%%%%%%%%%%%%%%%%

\end{document}